\documentclass[fleqn,twoside]{article}
\usepackage{espcrc2,psfig}

\usepackage{graphicx}
\usepackage[figuresright]{rotating}

\newcommand{\AmS}{{\protect\the\textfont2
  A\kern-.1667em\lower.5ex\hbox{M}\kern-.125emS}}

\title{Large Field Cutoffs Make Perturbative Series Converge}

\author{Yannick Meurice\address[MCSD]{Department of Physics and Astronomy, \\
The University of Iowa, \\
Iowa City, Iowa 52242, USA}
        \thanks{This research was supported in part by the Department of Energy
under Contract No. FG02-91ER40664.}}

\begin{document}

\begin{abstract}
For $\lambda \phi ^4$ problems, 
convergent perturbative series can be obtained 
by cutting off the large field configurations.
The modified series converge to values exponentially close to the exact ones.
For $\lambda$ larger than some critical value,
the method outperforms Pad\'e approximants and Borel summations.
We discuss some aspects of the  
semi-classical methods used to calculate the modified
Feynman rules and estimate the error associated with the procedure.
We provide a simple numerical example where the procedure works
despite the fact that the Borel sum has
singularities on the positive real axis. 
\vspace{1pc}
\end{abstract}

\maketitle
\section{INTRODUCTION}
It is intuitively clear that for Euclidean 
scalar field theory with quartic interactions, the suppression
${\rm e}^{-a^D\lambda \sum\phi_x ^4} $ makes the 
large field configurations unimportant.
This general idea has been tested explicitly by introducing sharp field cutoffs
in models which can be solved numerically. A first example is 
the spectrum of the anharmonic oscillator which can be calculated 
very accurately 
\cite{bacus} by using Sturm-Liouville theorem and  
requiring the wave function to vanish at 
some large value $\phi_{max}$ rather than
at infinity. A second example is the calculation \cite{gam3} 
of the zero-momentum correlation functions of Dyson's hierarchical model. 
This calculations requires the evaluation of the Fourier transform of 
the local 
measure and a sharp field cutoff can then be introduced at 
the first iteration, leaving 
the rest unchanged. 
In both examples, if the field cutoff is taken large enough the errors 
due to the cutoff can be made exponentially small. 

In addition to being a natural procedure in numerical calculations,
a large field cutoff cures the lack
of convergence of the perturbative series. 
It removes the instabilities appearing at negative $\lambda$. From 
a mathematical point of view, ${\rm e}^{-\lambda \phi ^4} $ has a Taylor
expansion which is
{\it uniformly} convergent over the disk $|\phi|\leq \phi_{max}$
and one can then interchange the sum and 
the integral. This statement has been checked numerically
in Ref. \cite{convpert} for the two 
cases mentioned
above. It was also shown that the modifications to the Feynman rules 
due to the field cutoff could
be calculated accurately by using semi-classical methods. In addition,
the errors made at a given order, as a function of 
$\lambda$ and the field cutoff, 
can be estimated without knowing the numerical
answer. At fixed $\lambda$, this allows an optimal choice of the field 
cutoff. Beyond a certain value of $\lambda$, this procedure outperforms 
Pad\'e approximants and the Pad\'e-Borel method. This is illustrated 
in Fig. \ref{fig:f3} and discussed
in Ref. \cite{convpert}. In the following, we first explain the general 
ideas with a simple example and then present some aspects not discussed 
in \cite{convpert}. 
\section{A SIMPLE EXAMPLE} 
In order to understand the origin of the divergence and how to remedy it,  
we first consider the simple integral
\begin{equation}
Z(\lambda)=\int_{-\infty}^{+\infty}d
\phi{\rm e}^{-(1/2)\phi^2-\lambda \phi^4}\ .
\label{eq:int}
\end{equation}
If we expand ${\rm e}^{-\lambda \phi^4}$, the integrand for the order $p$ 
contribution is ${\rm e}^{-(1/2)\phi^2}\phi^{4p}/p!$ and has its maximum when
$\phi^2=4p$. 
On the other hand, the truncation of ${\rm e}^{-\lambda \phi^4}$ at
order $p$ is accurate provided that $\lambda \phi^4 \ll p$. 
Requiring that the peak of the integrand for the $p$-th order term 
is within the range of values of $\phi$ for which the $p$-th order truncation
provides an accurate approximation, yields the condition $\lambda\ll 
(16p)^{-1}$.
One sees that the range of validity for $\lambda$ shrinks as one increases
the order. 
%as illustrated in Fig. \ref{fig:regpertint}. 
%Note that the same bound is obtained 
%by using the ``rule of thumb''.
We can avoid this problem  by restricting the range of integration in 
Eq. (\ref{eq:int}) to $|\phi|\leq \phi_{max}$. 
As the order 
increases, the peak of the
integrand moves across $\phi_{max}$ and the contribution is suppressed.
Similar conclusions were reached in Ref. \cite{pernice98} using 
Lebesgue's dominated convergence.
%It is easy to sho        

The coefficients of the modified series 
satisfy the bound $|a_p|<\sqrt{2\pi}\phi_{max}^{4p}/p!$ and 
the modified series
defines an entire function. However, we are now constructing a perturbative
series for a problem which is slightly different than the original one.
This procedure is justified from the fact that we have an exponential
control on the error $\delta Z$
due to the restricted 
range of integration:
\begin{equation}
%|Z(\lambda)-Z(\lambda , x_{max})|
\delta Z
<2{\rm e}^{-\lambda \phi_{max}^4}
\int_{\phi_{max}}^{\infty}d\phi{\rm e}^{-(1/2)\phi^2}\ . 
\label{eq:bound}
\end{equation}
%\begin{figure}[htb]
%\vspace{9pt}
%%\framebox[55mm]{\rule[-21mm]{0mm}{43mm}}
%\centerline{\psfig{figure=f1proc.EPS,height=1.8in}}
%\caption{
%Number of correct significant digits
%obtained with regular perturbation theory 
%at order 1, 2, 3, ..., 15 
%for the anharmonic oscillator, vs. log$_{10}\lambda$.
%As the order increases, the approximate lines
%rotate clockwise. The ``forbidden region'' is the upper right corner. }
%\label{fig:regpertint}
%\end{figure}
\section{SEMI-CLASSICAL ESTIMATES}
\begin{figure}[htb]
\centerline{\psfig{figure=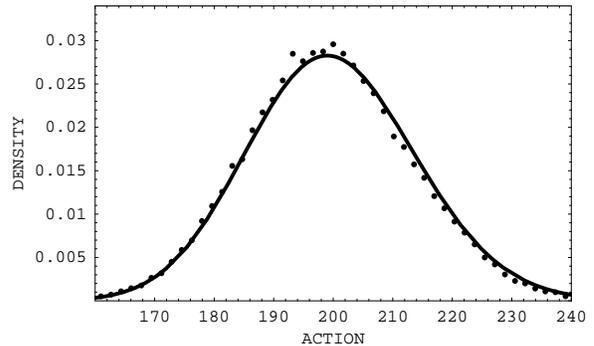,height=1.8in}}
\vskip-0.4in
\caption{Action histogram for the harmonic oscillator compared to 
$(\Gamma [N/2])^{-1/2}S^{N/2-1}{\rm e}^{-S}$ for $N=400$ sites.}
\label{fig:actdis}
\vskip-0.3in
\end{figure}

\begin{figure}[htb]
\centerline{\psfig{figure=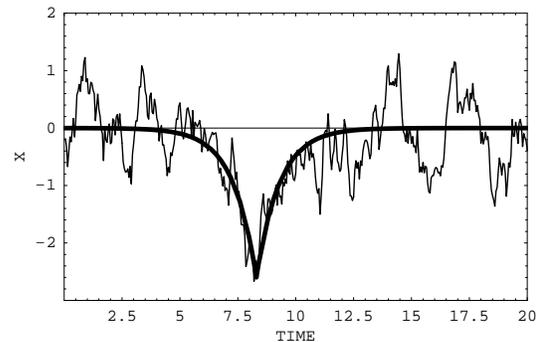,height=1.8in}}
\vskip-0.4in
\caption{A large field configuration (thin line) compared 
with $\phi_{max} {\rm e}
^{-|\tau-\tau_0|}$ (thick line) . }
\label{fig:expfit}
\vskip-0.2in
\end{figure}

Approximate  bounds of the form of Eq. (\ref{eq:bound}) 
for lattice models or their continuum limit 
can be obtained by semi-classical methods. The general idea is that
configurations with large field values are usually configurations of large
action and they can be described by classical solutions with small 
fluctuations. A detailed calculation has been performed in Ref. 
\cite{convpert} in the case of the anharmonic oscillator.
We used a dilute-gas approximation for configurations with 
one ``lump'' of large values to estimate the effect of the 
field cutoff on the ground state. In the continuum limit, at 
imaginary time and at $\lambda =0$, the relevant classical solutions are 
$\phi_{max} {\rm e}^{-|\tau-\tau_0|}$ (in $\hbar=m=\omega=1$ units). 
Such solutions
can be observed in MC simulations for the harmonic oscillator. 
We first checked that proper thermalization occurred by comparing
the distribution of the values of the action
with a simple analytical result for 
quadratic actions as shown in Fig. \ref{fig:actdis}. A typical 
configuration with large field values is shown in Fig. \ref{fig:expfit}
together with the classical solution.

Adapting standard semi-classical arguments, we obtained the 
zero-th order correction to 
the ground state: 
\begin{equation}
\delta E_0^{(0)} \simeq
4 \pi^{-1/2}\phi_{max}^{2}\int_{\phi_{max}}^{+\infty}d\phi{\rm e}^{-\phi^2}\ .
\end{equation}
The effects of a non-zero $\lambda$ are obtained by replacing  $\phi^4$ by 
classical solution at $\lambda=0$. One obtains the 
approximate error at order $n$:
\begin{equation}
|\delta E_o (\lambda)|\simeq\delta E_o^{(o)}
{\rm e}^{-(1/2)\lambda\phi_{max}^4}+
|a_{n+1}|\lambda^{n+1} \ . 
\label{eq:scbound}
\end{equation}
Both estimates are in very good agreement \cite{convpert} 
with numerical results provided that $\phi_{max}>2$.
\vskip-0.1in
\vskip-0.2in

\begin{figure}[htb]
\centerline{\psfig{figure=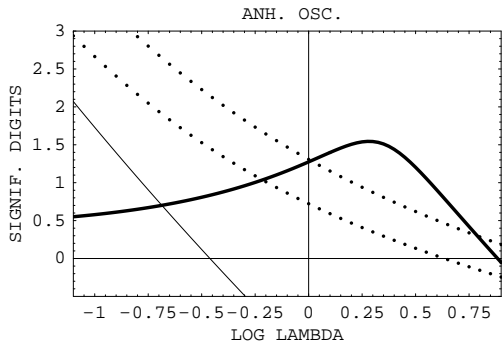,height=1.8in}}
\vskip-0.3in
\label{fig:f3}
\caption{Number of significant digits 
obtained with a large field cutoff (thick line), for an expansion at 
order 3, compared to regular perturbation (thin line), 
Pad\'e approximants for the regular 
series (bottom set of dots), and the 
Pad\'e-Borel method (top set of dots), at the same order.}
\vskip-0.3in
\end{figure}

The error due the field cutoff decreases when $\lambda$ increases, while
the error due to the truncation of the perturbative series at a finite
order decreases when $\lambda$ decreases. One thus reaches a compromise 
at intermediate values of $\lambda$ where the total error is minimized.
This is illustrated in Fig. \ref{fig:f3}.
By increasing the field cutoff, the optimal value of $\lambda$ decreases.
Using the approximate error formula, one can thus at the same time 
choose an optimal field cutoff and estimate the error.
\vskip-0.2in
\section{OTHER MODELS}
\begin{figure}[htb]
\centerline{\psfig{figure=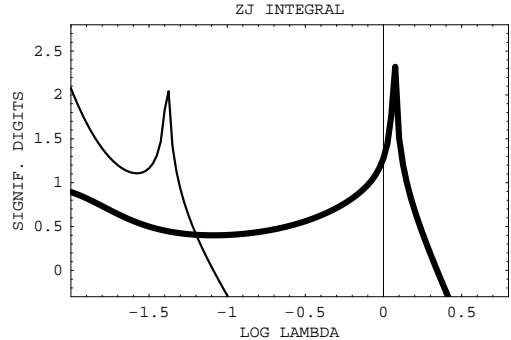,height=1.8in}}
\vskip-0.2in
\caption{Number of significant digits for an expansion of order 3 of 
Eq. (\ref{eq:zjint}) for $a=3/4$, with regular perturbation (thin line) 
and for a restricted range of $\phi$ (thick line).}
\label{fig:zjint}
\vskip-0.1in
\end{figure}

The generality of the basic principle invoked above 
(uniform convergence of the exponential on 
a compact neighborhood of the origin) implies that it 
can be used in many situations. In particular, for models 
where the Borel sums have singularities on the positive real axis.
A simple example is the integral \cite{zj}
\begin{equation}
\int_{-\infty}^{+\infty}d\phi {\rm e}^{-(1/2)\phi^2+((1+a)/3a)\sqrt{\lambda}\phi^3
-(1/4a)\lambda \phi^4}\ .
\label{eq:zjint}
\end{equation}
Numerical results for $a=3/4$ confirm that 
all the coefficients of the series are positive and that
the Pad\'e approximants of the series or of its Borel sums have 
all their poles on the positive real axis.
The series defined with cutoffs in $\phi$ give 
results which are significantly 
better than regular perturbation theory when $\lambda >0.1 $. 
This is illustrated in Fig. \ref{fig:zjint}. Complementary
discussions can be found in Refs. \cite{pernice98,bdj}.
\vskip-0.2in

In lattice gauge theory,
e.g. $SU(2)$ in 4 dimensions, 
$U_{x,\mu}={\rm e}^{iga{\bf A}_{\mu}.{\bf \tau}}$ with
${\bf A}_{\mu}$ contained in a sphere of radius ${\pi\over {ga}}$ for
each $\mu$ .
In the literature on lattice perturbation theory (with the exception 
of van Baal), one usually replaces $\int dg_{\mu}$ by 
$\int_{-\infty}^{+\infty}dA_{\mu}^i$ since
in the continuum limit the range becomes infinite.
From our point of view, the lattice spacing $a$ not only regulates the 
UV divergences but also the large order behavior of perturbation theory.
We plan to treat this problem by following a path similar to scalar field 
theory but with 
different approximations to treat 
the massless quadratic theory with a field cutoff.


\begin{thebibliography}{9}
%\bibitem{leguillou90}
%J.~C. Le Guillou and J. Zinn-Justin, {\em Large-Order Behavior of Perturbation
%  Theory} (North Holland, Amsterdam, 1990) ands Refs. therein.
\bibitem{bacus}
B. Bacus, Y. Meurice, and A. Soemadi, J. Phys. A {\bf 28},  L381  (1995).

\bibitem{gam3}
J. Godina, Y. Meurice, and M. Oktay, Phys. Rev. D {\bf 57},  R6581  (1998)
and D {\bf 59},  096002  (1999).

\bibitem{convpert}
Y. Meurice, University of Iowa preprint, hep-th/0103134.

\bibitem{pernice98}
S. Pernice and G. Oleaga, Phys. Rev. D {\bf 57},  1144  (1998).
%\bibitem{coleman}
%S. Coleman, {\em Aspects of Symmetry} (Cambridge University Press, Cambridge,
%  1985).
\bibitem{zj}
J. Zinn-Justin, {\em Quantum Field Theory and Critical Phenomena}, (Oxford, 
1989).
\bibitem{bdj}
I. Buckley, A. Duncan and H. Jones, Phys. Rev. D {\bf 47}, 2554 (1993).

\bibitem{vanbaal91}
P. van Baal, Nucl. Phys. B {\bf 351},  183  (1991).
\end{thebibliography}
\end{document}